\documentclass[manuscript]{aastex}
\usepackage{amsmath}

\slugcomment{}

\shorttitle{A quadruply lensed quasar in PS1}
\shortauthors{C. Berghea et al.}

\begin{document}

\title{ Discovery of the first quadruple gravitationally lensed quasar candidate with Pan-STARRS }

\author{C. T. Berghea\altaffilmark{1}, George J. Nelson\altaffilmark{1}, C. E. Rusu\altaffilmark{2}, C. R. Keeton\altaffilmark{3}, R. P. Dudik\altaffilmark{1}}

\altaffiltext{1}{U.S. Naval Observatory (USNO), 3450 Massachusetts Avenue NW, Washington, DC 20392, USA}
\altaffiltext{2}{Department of Physics, University of California, Davis, 1 Shields Avenue, CA 95616, USA}
\altaffiltext{3}{Department of Physics \& Astronomy, Rutgers, the State University of New Jersey, 136 Frelinghuysen Road, Piscataway, NJ 08854, USA}

\email{ciprian.t.berghea@navy.mil}

\begin{abstract}

We report the serendipitous discovery of the first gravitationally lensed quasar candidate from Pan-STARRS. The $grizy$ images reveal four point-like images with magnitudes between $14.9$ and $18.1$ mag. The colors of the point sources are similar, and they are more consistent with quasars than with stars or galaxies. The lensing galaxy is detected in the $izy$ bands, with an inferred photometric redshift of $\sim 0.6$, lower than that of the point sources. We successfully model the system with a singular isothermal ellipsoid with shear, using the relative positions of the five objects as constraints. While the brightness ranking of the point sources is consistent with that of the model, we find discrepancies between the model-predicted and observed fluxes, likely due to microlensing by stars and millilensing due to the dark matter substructure. In order to fully confirm the gravitational lens nature of this system, and add it to the small but growing number of the powerful probes of cosmology and astrophysics represented by quadruply lensed quasars, we require further  spectroscopy and high-resolution imaging.

\end{abstract}

\keywords{quasars: general  --- gravitational lensing: strong  --- cosmology: observations }

\section{INTRODUCTION}

Since the serendipitous discovery of the first gravitationally lensed quasar \citep{wal79}, these systems have become powerful probes of astrophysics and cosmology, as illustrated by the wealth of science they have provided over the years. For example, \citet{peng06} and \citet{din17} have studied the coevolution of supermassive black holes and the quasar hosts harboring them up to high redshift \citep{cla02,tre10}. \citet{ogu12}, \citet{bon17} and others have constrained the cosmological constant and the Hubble constant from samples of lensed quasars \citep[see the recent review by][]{Tre16}. Flux ratio anomalies have revealed luminous satellites or set constraints on the dark matter substructure in the lensing galaxies \citep[e.g.,][]{chiba05,mck07,fad12}. Ensembles of lenses have revealed the structure of massive galaxies \citep[e.g.,][]{koc00,ogu14}. Other insights into quasar accretion disks \citep[e.g.,][]{dai10}, broad-line regions \citep[e.g.,][]{slu12}, and black hole spin \citep{reis14} have been gained.

These applications all require increasingly large samples of lensed quasars, and in particular, quadruply lensed quasars (quads), due to the increased number of modeling constraints they provide. To date, there are only about three dozen known quads over the whole sky\footnote{http://masterlens.astro.utah.edu/}. Due to the rare nature of lensed quasars, large-scale surveys of sufficient depth and resolution are required to significantly increase the present sample \citep{ogu10}. As such surveys have become available, pioneering searches for lensed quasars, such as the Cosmic Lens All Sky Survey \citep[CLASS;][]{mye03} have been succeeded by the SDSS Quasar Lens Search \citep[SQLS;][]{ogu06}, and more recently by the STRong lensing Insight in the Dark Energy Survey \citep[STRIDES;][]{agn15} and the extension to SQLS using the Baryon Oscillation Spectroscopic Survey \citep[BQLS;][]{more16}, amongst others.

In this paper, we report the serendipitous discovery of a quad lensed quasar candidate from the Panoramic Survey Telescope and Rapid Response System (Pan-STARRS1, hereafter PS1) released images. The lensed quasar was discovered as part of a variability study of active galactic nuclei selected based on Mid-Infrared characteristics \citep{sec01}, and data from the USNO Robotic Astrometric Telescope \citep[URAT;][]{zac15}. To our knowledge, if proven with spectroscopic observations, this would be the first published gravitational lens discovered in the PS1 data. We note that this system is very similar to the well-studied quad lens RX J1131-1231 \citep{slu03}, in terms of image separation and overall configuration. We model the system to provide evidence in support of its lensing nature, and we conclude with the necessity of obtaining spectroscopic confirmation. Due to its position, from the ground this is not possible until the end of the year.

PS1 is a wide-field imaging system, with a 1.8 m telescope and 7.7 deg$^2$ field of view (FOV), located on the summit of Haleakala in the Hawaiian island of Maui. The first PS1 data was released in 2016 December, including both images and catalogs \citep[see][]{cha16}. The 1.4 Gpixel camera consists of 60 CCDs with pixel a size of 0.256 arcsec \citep{ona08, ton09}. It uses five filters (g$_{P1}$, r$_{P1}$, i$_{P1}$, z$_{P1}$, y$_{P1}$, hereafter $grizy$), similar to the ones used by the Sloan Digital Sky Survey \citep[SDSS;][]{york00}. The largest survey PS1 performs is the 3$\pi$ survey, covering the entire sky north of $-30\deg$ decl. Given the large sky coverage, resolution, and depth, PS1 is expected to contain nearly 2000 gravitationally lensed quasars, of which about 300 will be quad lensed quasars \citep{ogu10}.
 
The structure of the paper is as follows.
In Section 2, we measure the relative astrometry, photometry, and morphology of this system. In Section 3, we infer photometric redshifts, and in Section 4, we conduct a photometric variability study to support the photometric redshifts. In Section 5, we fit a lensing mass model to the system. We present our conclusions and future works in Section 6. We assume a cosmological model with $\Omega_M = 0.274$, $\Omega_L = 0.726$, and $h$ = 0.71. All magnitudes are in the AB system except for the WISE ones, which are in the Vega system. 

\section{ASTROMETRY, PHOTOMETRY, AND MORPHOLOGY}

\begin{figure}
\epsscale{1.0}
\plotone{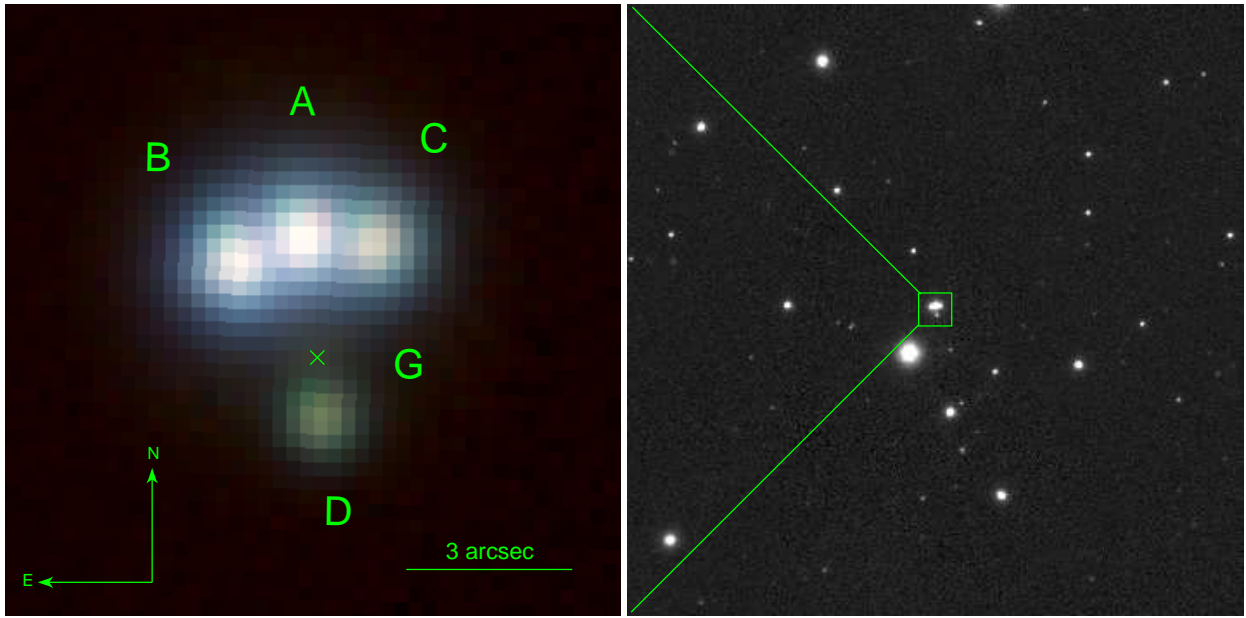}
\caption{
PS1 images of the lens candidate showing the four quasar images (A-D) with the lensing galaxy position G marked with an x. Left: close-up color image using the $g$ (blue), $i$ (green), and $y$ (red) filters. The {\it Gaia} position for component D (26.792307, 46.511273) is used as absolute astrometric reference position for this system. Right: $y$-band image of a 2 $\arcmin$ region around the system. The bright star southeast of the lens is saturated in the other bands.}
\label{lens}
\end{figure}

We used the stacked images from PS1 in all five filters in order to perform a morphological modeling. We present a close-up color image of the system in Fig.~\ref{lens}, which clearly shows four objects within $\lesssim 3.8\arcsec$ of each other. The three brightest of these (A, B and C) are arranged in an arc-like configuration. The PS1 catalog identifies only the sources A and D, with B and C being blended with A. 

\begin{figure}
\epsscale{1.0}
\plotone{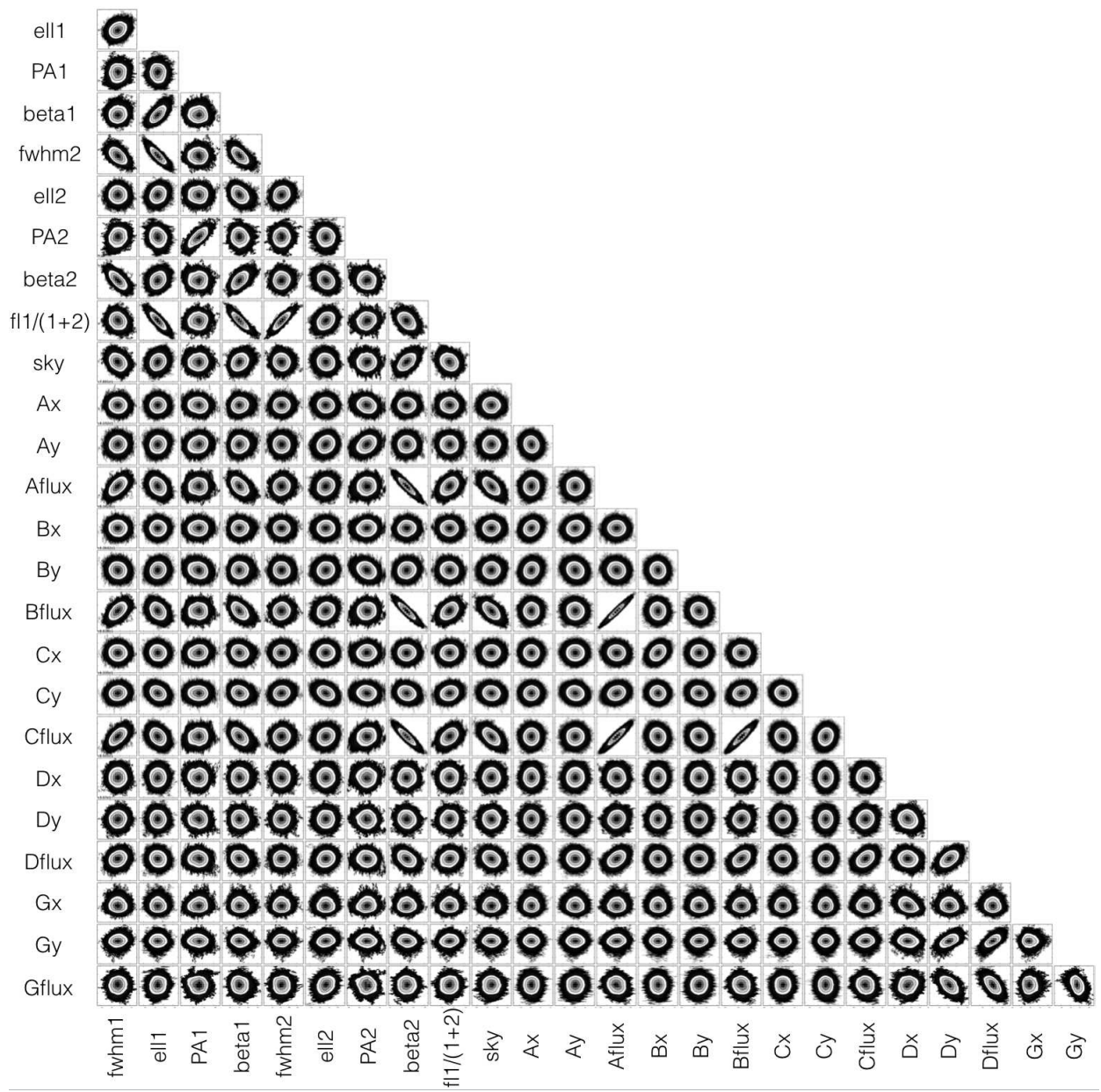}
\caption{
Covariances between the parameters used to model the observed system in $i$-band with Hostlens. The three contour lines mark the 1-, 2- and 3-$\sigma$ limits. The parameters are, in order, the FWHM in arcseconds, ellipticity, orientation, the shape of the first Moffat profile (1,2,3,4), the same for the second Moffat profile (5,6,7,8), the relative flux as flux1/(flux1 + flux2) (9), sky level in counts (10), positions in pixels on the x and y axes of point source A (11,12) and its flux in counts (13), and the same for objects B,C and D (14-25).}
\label{mcmc_hostlens}
\end{figure}

\begin{figure}
\epsscale{1}
\plotone{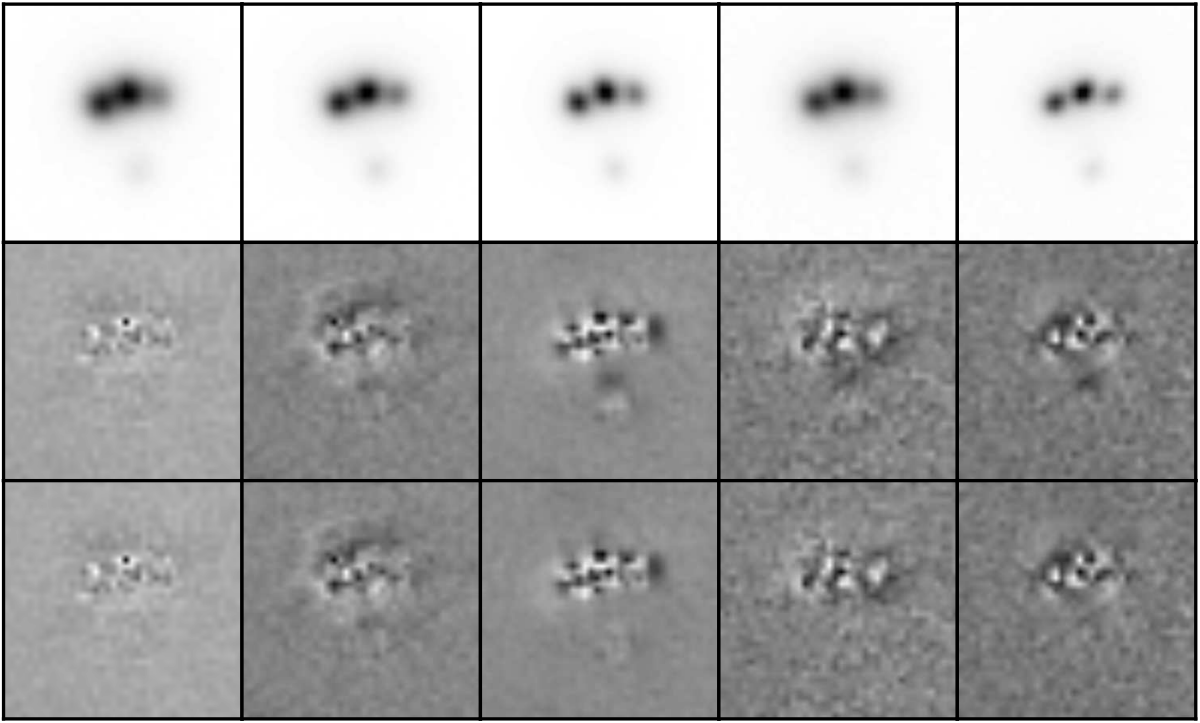}
\caption{
Original images ($10\arcsec\times10\arcsec$, first row) and residuals after subtracting the best-fit model with Hostlens. From left to right: $g$, $r$, $i$, $z$, and $y$ bands. In the second row, the lensing galaxy is not included in the modeling, whereas in the third row it is. The images are in linear scale and cover the full dynamic range. The remaining structure in the residuals amounts to less than 6\% of the peak fluxes and can be attributed to the large Poisson noise at the center of the point-like images, as well as to the faint arcs expected from the lensed quasar host galaxy, which are not accounted for in the modeling.}
\label{hostlens}
\end{figure}

The simple modeling with a point spread function (PSF) constructed from stars in the PS1 FOV left large residuals at the locations of the four objects, which were significant enough to affect the derived parameters. We suspect that this is due to spatial variation in the PSF across the PS1 FOV, and have therefore adopted a technique of fitting the system with an analytical PSF, which, it is assumed will not change on the small scale represented by the system. We did this using Hostlens \citep{rus16}, a variant of glafic \citep{ogu10b} that uses $\chi^2$ minimization in order to fit point sources as well as S{\'e}rsic \citep{ser63} profiles convolved with an analytical PSF. The analytical PSF comprises two concentric \citet{mof69} profiles, each one with its own FWHM, ellipticity, orientation, and shape parameter. The two profiles are also characterized by the relative flux of one to the other. We successfully modeled the four objects as point sources convolved with the PSF, therefore showing that they are point like. Our strategy was to run Hostlens starting from 100 different positions in the parameter space, select the best of the resulting models, and further run 10 Markov Chain Monte Carlo (MCMC) chains using the Metropolis-Hastings algorithm around it. The chains consist of one million steps, with an acceptance rate of $\sim0.3$, and we removed the first one-fifth of these (the ``burn-in'' steps). We also checked that the chains have converged, using the method of \citet{gel95}. The MCMC-derived uncertainties between the various analytical parameters are shown in Fig.~\ref{mcmc_hostlens}, and generated using corner.py \citep{for16}. In order to quantify the additional uncertainties inherent in our method due to the non-analytical nature of the ``true'' PSF, we ran 100 simulations where we added noise of similar properties to the real images on top of the best-fit model. Here, we created the best-fit model by using a PSF constructed from nearby stars, and we ran Hostlens on each of the 100 simulations, using an analytical PSF. As a final estimate of our uncertainties, we used the maximum between the MCMC-derived uncertainties and those from the simulations. We were able to obtain much improved residuals in all bands, and we show these in Fig.~\ref{hostlens}. 
 
Fig.~\ref{hostlens} shows conspicuous residuals in the $i$, $z$ and $y$ bands (middle row). These are consistent with the discovery of a lensing galaxy, in support of the lensing nature of this system. Due to the relatively low signal-to-noise ratio and the proximity of the point sources, we were unable to fit the morphology of this galaxy with free parameters. The red colors, suggested by the non-detection in the bluer bands, suggest that it is a red, early-type galaxy. As a result, we modeled it with a S{\'e}rsic index of 4, typical of early-type galaxies. We checked that this produces a better fit than a S{\'e}rsic index of 1. For the S{\'e}rsic profile, we used an axis ratio of 1 and we fixed the effective radius at $0.5\arcsec$. As the galaxy is most conspicuous in the $i$ and $y$ bands, we consider the most reliable estimate of the relative positions of all objects to be derived from the weighted average of the positions measured in these two bands, where we weight by the inverse of the measured uncertainties. The resulting astrometry and photometry in each band are shown in Table~1. We have corrected the magnitudes for galactic extinction using the maps of \citet{sch11}.

In addition to the PS1 data, we have also looked for archival data of this system. The system is detected in both the {\it Gaia} \citep{gaia} DR1 and PS1 catalogs; however, objects A, B, and C are blended into a single object by the automatic pipeline. PS1 psf magnitudes for A are in general smaller than what we obtained, probably due to blending, but for D they are very similar to our uncorrected magnitudes: 18.599$\pm$0.013,  18.153$\pm$0.002,  18.079$\pm$0.007,  17.843$\pm$0.0093,  17.534$\pm$0.023 in the $grizy$ bands, respectively. We use the {\it Gaia} position for component D as the absolute astrometric reference position for this system: 26.792307, 46.511273. The errors for these coordinates given in the {\it Gaia} catalog are 15.6 and 8.9 mas, respectively.

%%PS1 psf magnitudes for A are in general smaller than what we obtained probably due to blending, but for D they are very similar to our uncorrected magnitudes: 18.599$\pm$0.013,  18.153$\pm$0.002,  18.079$\pm$0.007,  17.843$\pm$0.0093,  17.534$\pm$0.023. Gaia G magnitudes are 15.17 and 18.13 for A and D, respectively.

The system is also bright in the infrared, with the unresolved magnitudes being 11.524$\pm$0.022, 10.434$\pm$0.020, 6.769$\pm$0.015, and 4.518$\pm$0.023 in the WISE bands $-$ W1, W2, W3, and W4, respectively. 
%%The WISE colors W1$-$W2$=$1.1 and W2$-$W3$=$3.7 match very well those of quasars for redshift below 3.5 \citep[e.g.][]{wu12}. 

Finally, the system is very likely also a radio source. The NRAO VLA Sky Survey (NVSS) images show a source, 2MASX J01471020+4630433, with flux 12.6 mJy at 1.4 GHz. The listed position for this source is only $3\arcsec$ away from the center of the system in the optical bands, which is comparable with NVSS astrometric errors \citep{con98}. We are currently in the process of further observing the system in the radio with the Very Large Array (VLA). Radio data are useful for the study of lensed quasars, as it provides high-resolution imaging of the source, as well as fluxes not affected by microlensing and extinction.

\section{PHOTOMETRIC REDSHIFTS}

\begin{figure}
\epsscale{1.1}
\plottwo{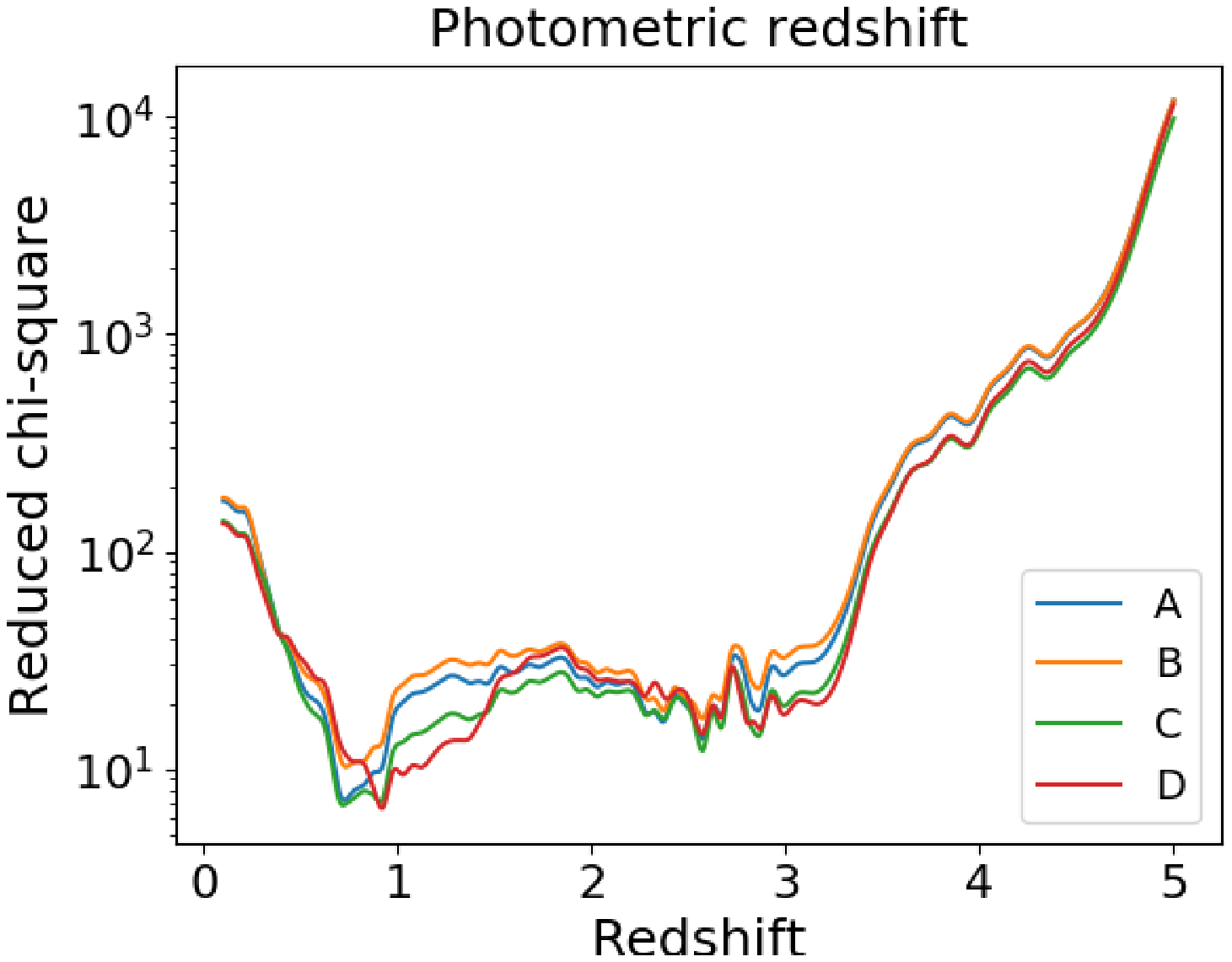}{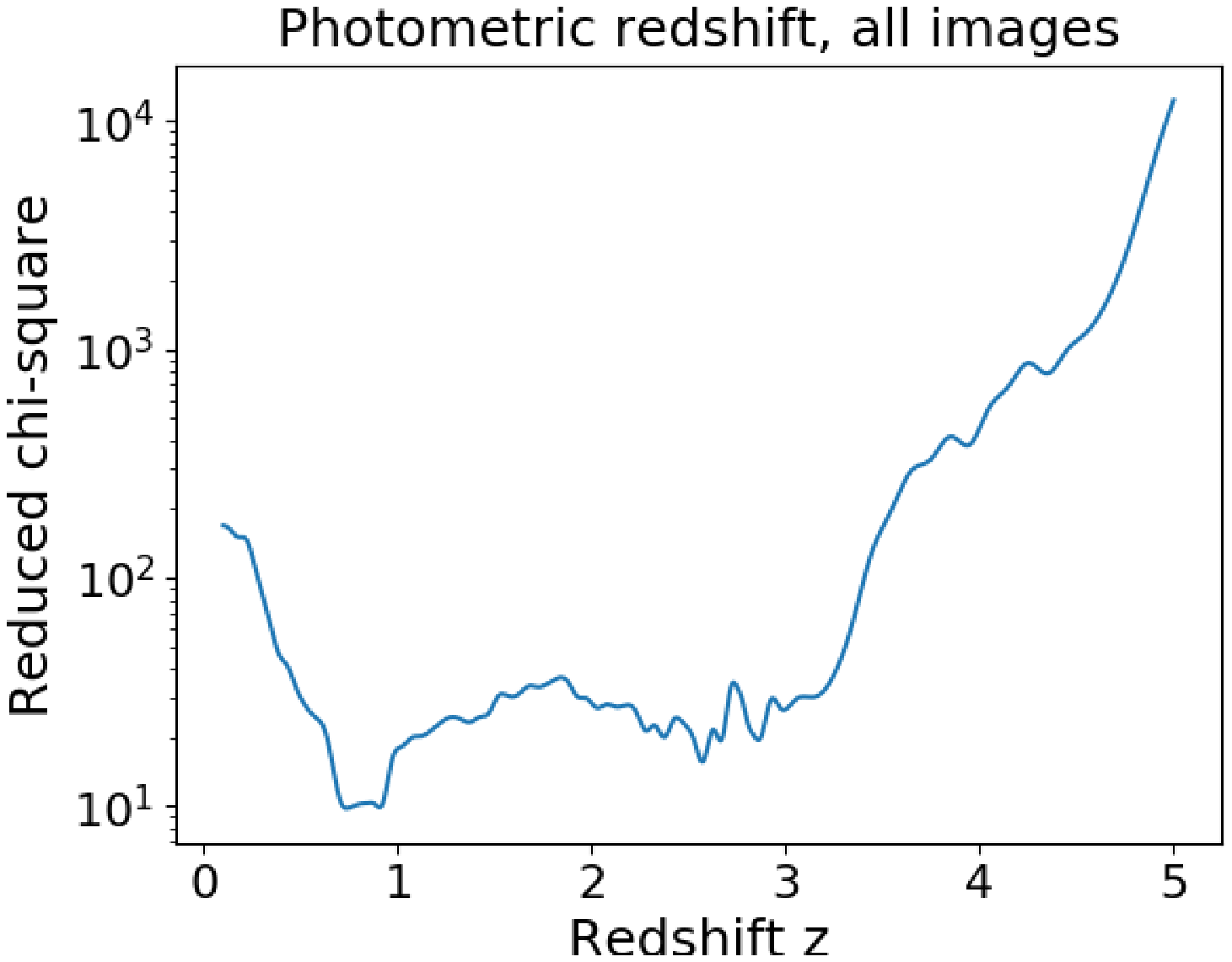}
\caption{
Photometric redshift estimates using the method in \citet{wu10}. 
Left: individually for each image; right: for all images simultaneously.}
\label{redshift}
\end{figure}

We fit the photometry measured in Section 2 for the point-like sources and the faint red galaxy using spectral energy distribution (SED) templates. The colors of the point-like images are very similar (Table~1), consistent with being multiple images of the same source. Due to their point-like morphology, we suspect these to be quasars, and have therefore fit them with quasar templates in order to check the quality of the fit and to infer photometric redshifts. Photometric redshifts of quasars are notoriously difficult to constrain, due to the relatively featureless continua \citep{ric09} and the variable equivalent widths of the broad emission lines. We measured redshifts using several methods:
\begin{itemize}
\item The method of \citet{wu10} uses derived quasar colors as a function of redshifts. This is based on SDSS colors. PS1 bands are similar but not identical to the SDSS bands, and we used the corrections of \citet{fin16} to obtain magnitudes in the SDSS system. Following \citet{wu10}, we minimized ${\chi}^2$ to obtain redshifts for each of the point sources. We only used three of the colors as the Y band from UKIRT is very different from the PS1 y band. We plot $\chi^2_{\nu}$ in Figure~\ref{redshift} for each quasar image, and also by fitting all sources together. We notice that the curves are quite consistent with each other, as expected from the similar colors of the sources. In the latter case, we obtain the best fit at $z = 0.820^{+0.018}_{-0.014}$. We notice a second minimum at $z \sim2.6$. The $\chi^2_{\nu}$ for the minima is large, which could be due to the large spread seen in the colors of \citet{wu10}. We note that no numerical uncertainties are listed in that paper for the derived colors. We have also explored incorporating a prior based on the quasar luminosity function, given the unmagnified observed magnitudes of the source (see Section 5). While the prior prefers low-redshift values, its effect on Fig.~\ref{redshift} is negligible.
\item A second method uses the photometric redshift code LePhare \citep{arn99,ilb06} to fit a quasar template to the observed colors. We obtain the best fit at a redshift of $\sim 2.6-2.8$ for all four point sources. These results agree with the second peak of the method above. The $\chi^2$ is between 13 and 125 for 4 degrees of freedom (5 filters). We find poorer fits (larger $\chi^2$) when using stellar or galactic SED templates, providing further support that the sources are quasars. In particular, the best-fit galaxy templates correspond to starburst galaxies at low redshift $\lesssim0.1$. At such low redshifts the sources would likely be resolved, not point like. 
\item A combination of SDSS and WISE photometry provides good discriminators between high- and low-redshift quasars. \citet{pom15} used the ratio between WISE W1 and W2 bands to the $i$-band flux, and found that high values are expected for quasars below a redshift of 1. We added the fluxes for A, B, C, and D in the $i$ band (to account for the fact that the system is not resolved in WISE; G is of negligible flux), and we obtained a total magnitude of 14.4. We obtain ratios W1/$i$ = 14.4 and W2/$i$ = 39.0, which both indicate a redshift $>$ 2, and are not compatible with the low value of $\sim 0.8$, which would require ratios of W1/$i$ = 50 and W2/$i$ = 120. We also used the diagnostics from \citet{wu12}. As seen in Fig.~13 of that paper, the colors $z$ - W1 and W2 - W3 discriminate well between low and high redshift. We obtain $z$ - W1 = 2.8 and W2 - W3 = 3.7, which both suggest a redshift larger than 2. Our low value of 0.8 would require z - W1 = 4.0 and W2 - W3 = 3.0.

\end{itemize}

The lens galaxy G is very faint but we managed to fit a galaxy SED with an early-type template, resulting in a best-fit $z = 0.57^{+0.20}_{-0.13}$. For this, we used the magnitudes measured in the $izy$ bands and assumed detection limits of 21 mag in the $gr$ bands, $\sim$ 2 mag fainter than the rest. We independently checked that the observed magnitudes are reasonable for a redshift of $z=0.57$: for this redshift, the observed magnitudes imply an absolute magnitude of about $-23.0 \sim -23.5$ in the $R$-band. From the Faber-Jackson relation \citep{fab76}, using the velocity dispersion we estimate from the lens modeling in Section 5, we obtain $R \sim -23.6$, therefore not inconsistent.

\section{URAT VARIABILITY}

Here we undertake a quasar variability study as an independent check of the redshift estimate we derived in the previous section.
URAT1 was a U. S. Naval Observatory project designed to provide accurate astrometry in the northern hemisphere down to about 18 mag. A single filter centered at $\sim$~720 nm was used, which is almost as red as the PS1 $i$ filter. Observations at the Flagstaff Station over more than 3 years provided more than 50 epochs for a large number of objects. 
%The gravitational lens was discovered as part of project to study the variability of a sample of AGN selected based on WISE colors \citep{sec01}. We now present the results of this study for our quad lens, keeping in mind that the system was not resolved by URAT. 

To characterize the variability of this system, we used a typical method used for quasars, the structure function V, as presented in \citet{van04}. It uses the magnitude differences between different epochs of observation to estimate the variability at different time intervals (time lag). 
It is defined as
\begin{eqnarray}
  V = \left(\frac{\pi}{2} {\langle \vert \Delta m \vert \rangle}^{2}
      - \langle \sigma^{2} \rangle\right)^{\frac{1}{2}}\phn,
\end{eqnarray}
where $\Delta m$ is the measured magnitude difference for each pair of epochs in the light curve, $\sigma$
is the statistical measurement uncertainty of $\Delta m$  and the brackets denote average
quantities.

\citet{van04} used the structure function to characterize the variability of a large sample of SDSS quasars. We note that their study is statistical in nature and therefore only provides general properties which we use to try to discriminate between the two photometric redshifts we obtained in the previous section. 

In order to improve the results, we performed relative photometry of URAT epoch data using nearby stars in the 12 to 14 magnitude range. We removed stars that show variability and also removed some poor quality observations. We used more than 50 stars and 36 epochs spanning over two years. We estimate errors using the nearby stars with similar magnitudes. We obtained V = 0.074 and an average time lag of 297 days. According to Fig.~16 of \citet{van04}, this V value matches quasars at about $z = 2$ and therefore our photometric value of $z = 2.6$ is a better match than the low value of $z = 0.8$, which would require a V value of 0.05. We also used the absolute magnitude dependence in Fig.~11 of \citet{van04}. We estimate an absolute magnitude for our source quasar using the demagnified average of the observed magnitudes of A, B, and C (see Section 5). Assuming a composite quasar spectrum, we estimate that the $i$-band rest-frame absolute magnitude ranges between $-24.8$ and $-25.2$ at redshift 0.8, and between $-27.6$ and $-27.9$ at redshift 2.6. Again, we see that the V value better matches a high luminosity quasar, and therefore the higher redshift is preferred. We note that in this study we have ignored the variability caused by microlensing (see the next section).

%%The restframe time lag is 165 and 83 days for the assumed redshifts of 0.8 and 2.6, respectively. The restframe wavelength is 400 and 200 nm for the two cases. While Fig.~8 in \citet{van04} shows that for the I band a value of V = 0.073 matches a restframe timelag of about 30 days, the smaller value of 83 days corresponding to z = 2.6 provides a better match than the longer timelag of 165 days, which corresponds to V $>$ 0.1

\section{LENS MODELING}

The results of the previous sections show that the source is likely a quadruply imaged quasar, lensed by a foreground early-type galaxy. We subsequently modeled the observed configuration (the relative positions of the objects A,B,C,D, and G) using the lensmodel software from the gravlens package \citep{kee01}. We chose not to use the observed relative fluxes as constraints, as it is well known that at these wavelengths they usually show large discrepancies due to microlensing, millilensing, intrinsic variability, or extinction \citep[e.g.,][]{slu08}. For the mass profile of the lensing galaxy, we use a Singular Isothermal Ellipsoid (SIE) with external shear. There are eight constraints from the four image positions and two constraints from the galaxy position, for a total of 10. Also, there are nine free parameters: two for the galaxy position, one for the Einstein radius, two for ellipticity, two for shear, and two for the source position. The model therefore has one degree of freedom.

\begin{figure}
\epsscale{1.1}
\plotone{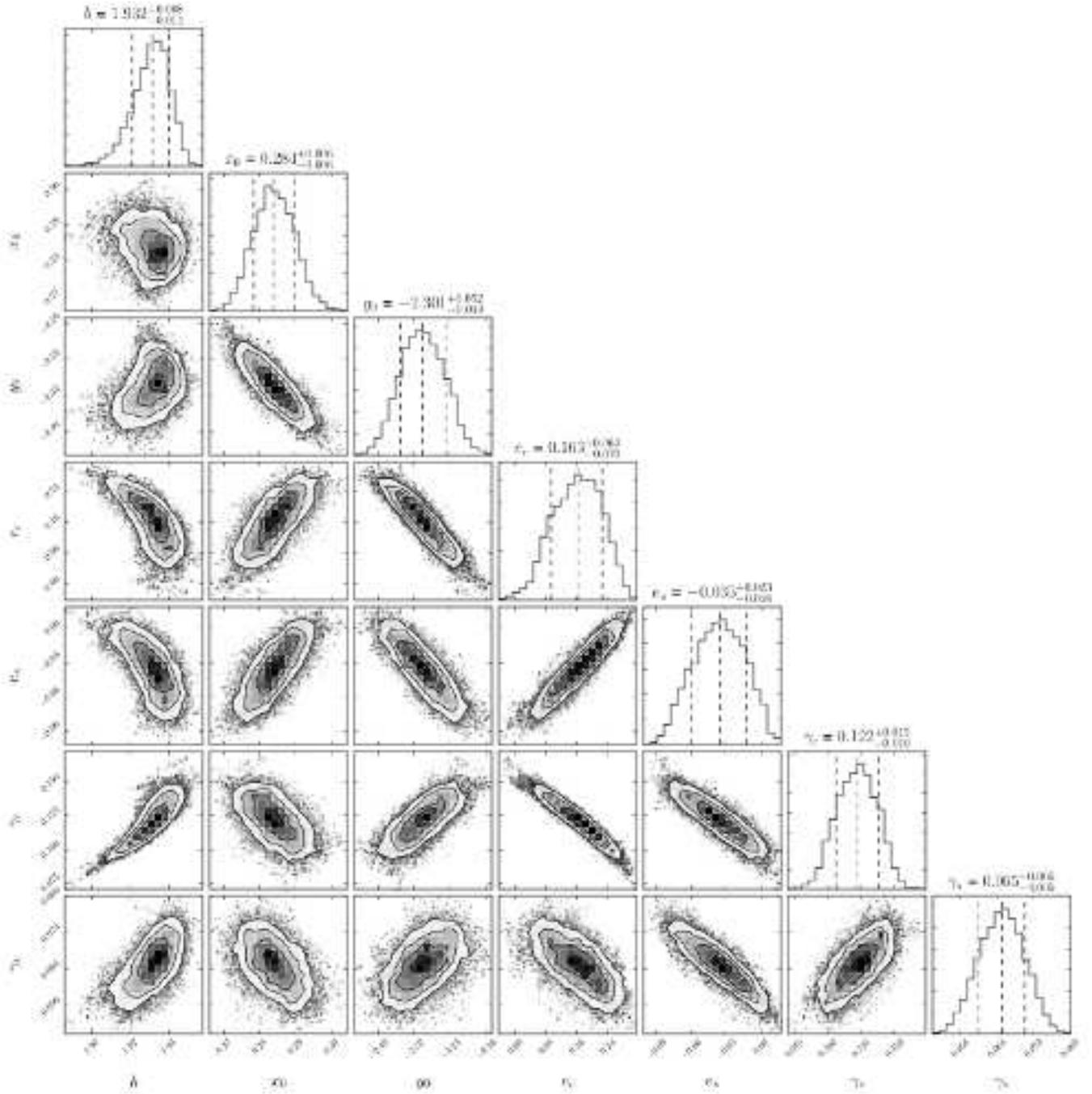}
\caption{
Covariances between the parameters used to model the observed system with lensmodel. The three contour lines mark the 1-, 2- and 3-$\sigma$ limits. The vertical lines in the histograms mark the 16th, 50th, and 84th percentiles. The parameters are, in order, the Einstein radius, lensing galaxy x and y positions, two paramaters for ellipticity, and two for shear.}
\label{mcmc_gravlens}
\end{figure}

For the best-fit mass model parameters, we find an Einstein radius $b = 1.932_{-0.011}^{+0.008}$, ellipticity $e_c = 0.163_{-0.070}^{+0.063}$, $e_s = -0.035_{-0.024}^{+0.023}$, and shear $\gamma_c = 0.122_{-0.016}^{+0.015}$, $\gamma_s = 0.065\pm0.005$. Here, we expressed the ellipticity and shear in quasi-Cartesian coordinates, rather than polar coordinates, as the distribution of uncertainties is more gaussian. The two coordinate systems are related by $(e_c,\gamma_c) = (e,\gamma)\cos2(\theta_e,\theta_\gamma)$ and $(e_s,\gamma_s) = (e,\gamma)\sin2(\theta_e,\theta_\gamma)$. The derived image magnifications are approximately 19, 11, 11, and 0.7 for A, B, C and D, respectively. The uncertainties were determined using MCMC as plotted in Fig.~\ref{mcmc_gravlens}, where the convergence of the chains was again checked using the method of \citet{gel95}. The $\chi^2$ for this model is 0.8. 

While the parameters derived above are independent of redshift, estimating the expected time delays between the images does require redshift information. Using the results of Section 3, we have fixed the redshift of the source at 0.57. If we assume a source redshift of 0.82, according to the best-fit result of the \citet{wu10} method in Section 3, we derive from the expression of the Einstein radius in a singular isothermal profile, $b=4\pi\left(\frac{\sigma}{c}\right)^2\frac{D_{ls}}{D_{s}}$, a velocity dispersion of $\sigma\sim500$ km/s. Here, $c$ is the speed of light, and $D_{s},D_{ls}$ are the angular diameter distances to the source and between the lens and source, respectively. This value is too large for a single galaxy, but if we employ the second-best source redshift estimate, $z\sim2.6$, which is preferred by LePhare, the combination of WISE and PS1 colors (Section 3) and the variability results in Section 4, we obtain $\sigma\sim315$ km/s, a more common value for massive early-type galaxies. We therefore assume this redshift to infer the predicted time delays, and we obtain $\Delta t_{BC} = 0.1$ days, $\Delta t_{AC} = 1.7$ days and $\Delta t_{DC} = 226.2$ days, with image C (a minimum of the time delay surface) leading. Image B is another minimum, and images A and D are saddle points.   

\begin{figure}
\epsscale{0.5}
\plotone{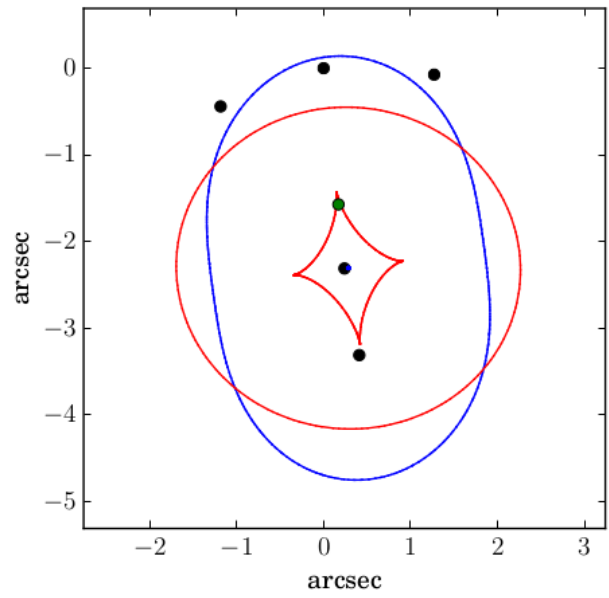}
\caption{
Gravitational lens modeling. The blue and red lines show the critical curves and caustics, respectively. The green dot shows the quasar position in the source plane, and the black dots show the lens and the observed image positions.}
\label{model}
\end{figure}

\begin{figure}
\epsscale{1.0}
\plotone{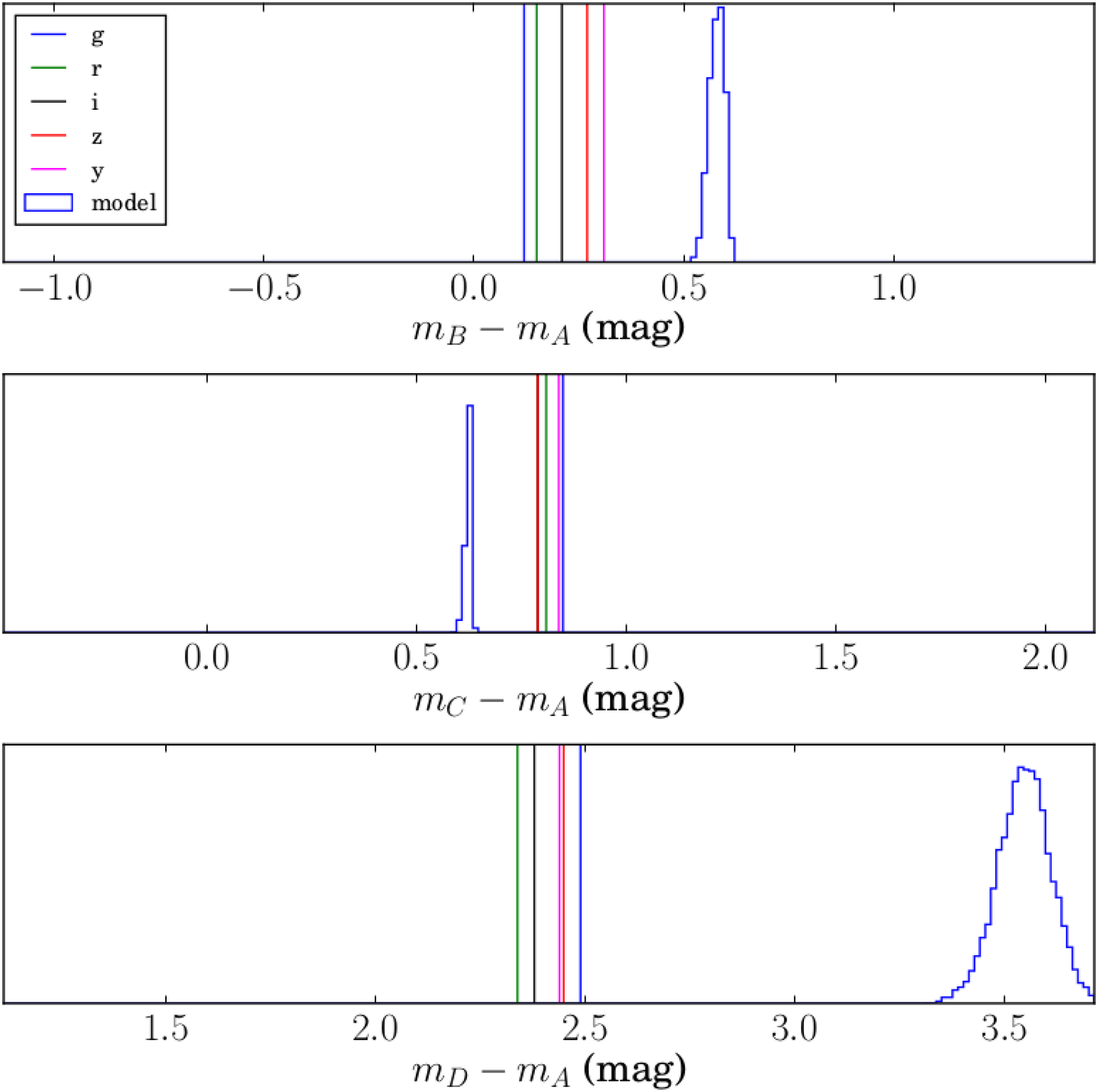}
\caption{
Comparison between the observed quasar image magnitudes relative to image A (vertical colored lines) and the model predictions (blue histogram) for each PS1 band. The model predictions were drawn from an MCMC exploration of the range of models.}
\label{modelphot}
\end{figure}

In Figure~\ref{model} we show the critical curve and the caustics, as well as the positions of the source and the observed images. The source is crossing a cusp, which results in the production of three bright images in close proximity and one isolated faint image. This is a classic configuration for quad lenses. The predicted magnitude differences are shown in Table~1, and a comparison with the observed values are shown in Figure~\ref{modelphot}. 

We find that the predicted brightness ranking of the images is consistent with the observations, with image A being brightest and image D being faintest. However, there are still large discrepancies between the observed and predicted fluxes. For example, the fluxes of images B and C are different, whereas they are expected to be the same for any smooth mass profile of the lensing galaxy. Such ``flux anomalies'' have been widely encountered for cusp lensed quasars and can be explained by small-scale structure in the lensing galaxy \citep[e.g.,][]{kee03}. We also remark that the colors of the four images are very similar, within $\lesssim 0.2$ mag across all filters. Small chromatic variations between multiple quasar images are known to be caused by intrinsic variability, microlensing, or differential extinction \citep{yon08}. In any case, differential extinction would not explain the flux discrepancy in image D, as this is $\sim1$ mag \emph{brighter} than expected. Since image D is closest to the lensing galaxy, the stellar density is comparatively high at its location, meaning that microlensing due to stars in the lensing galaxy is a plausible explanation for the discrepancy. Second, we remark that the time delays between image D and the other images is large, which means that while intrinsic flux variations between A, B and C will be washed out, they could stand out in image D, and therefore be responsible for the discrepancy. Lastly, small-scale substructure (millilensing) could cause contribute to the discrepancy.

\section{CONCLUSIONS AND FUTURE WORK}

We have presented the first quadruply imaged gravitationally lensed quasar candidate in the Pan-STARRS1 Survey, discovered via visual inspection of the multiband images. We find that the evidence supporting the gravitational lens nature of this system is overwhelming and consists of the following:

\begin{itemize}
\item The presence of four point-like images within $\lesssim 3.8\arcsec$ of each other, with similar colors, more consistent with quasar templates, but less so with stars and galaxies.
\item The detection of a red galaxy between the point source, such that the relative positions of all five objects are fully consistent with a well-known ``cusp'' configuration. We successfully reproduce this configuration using a SIE + shear model for the lensing galaxy. Furthermore, the brightness ranking of the point-like images is consistent with the one predicted by the model.
 \item The photometric redshift analysis shows that the galaxy is of elliptical template, typical for lensing galaxies, and of redshift smaller than the point sources. For the inferred redshifts (assuming a source redshift of $\sim2.6$), the velocity dispersion of the lensing galaxy is consistent with those of massive early-type galaxies.
\end{itemize}

Nevertheless, we stop short of claiming that this is a confirmed lens until we can obtain spectroscopy of this system. Spectroscopy of the four point-like images would unequivocally demonstrate whether they are multiple images of the same background quasar, as well as allow one to infer its redshift and physical properties. We also note that the system is accessible to ground-based high-resolution observations at most adaptive optics-capable facilities in the northern hemisphere, due to the its proximity to a $R\sim12$ mag star $\sim18\arcsec$ away.

In the introduction, we noted its similarity with RX J1131-1231; however our system appears to have a larger source redshift and images brighter  by about 2 mag \citep{slu06}. \citet{slu08} also measure flux discrepancies for RX J1131-1231, and show that these flux discrepancies can be explained by microlensing. 

\acknowledgments

We would like to thank Arunav Kundu, Shobita Satyapal, Nathan Secrest, Bryan Dorland, and Valeri Makarov for useful discussions and help. We also thank the referee for great suggestions. C.E.R was funded through the NSF grant AST-1312329, ``Collaborative Research: Accurate cosmology with strong gravitational lens time delays'', and the HST grant GO-12889. 

The Pan-STARRS1 Surveys (PS1) and the PS1 public science archive have been made possible through contributions by the Institute for Astronomy, the University of Hawaii, the Pan-STARRS Project Office, the Max-Planck Society and its participating institutes, the Max Planck Institute for Astronomy, Heidelberg and the Max Planck Institute for Extraterrestrial Physics, Garching, The Johns Hopkins University, Durham University, the University of Edinburgh, the Queen's University Belfast, the Harvard-Smithsonian Center for Astrophysics, the Las Cumbres Observatory Global Telescope Network Incorporated, the National Central University of Taiwan, the Space Telescope Science Institute, the National Aeronautics and Space Administration under Grant No. NNX08AR22G issued through the Planetary Science Division of the NASA Science Mission Directorate, the National Science Foundation Grant No. AST-1238877, the University of Maryland, Eotvos Lorand University (ELTE), the Los Alamos National Laboratory, and the Gordon and Betty Moore Foundation.

This work has made use of data from the European Space Agency (ESA)
mission {\it Gaia} (\url{https://www.cosmos.esa.int/gaia}), processed by
the {\it Gaia} Data Processing and Analysis Consortium (DPAC,
\url{https://www.cosmos.esa.int/web/gaia/dpac/consortium}). Funding
for the DPAC has been provided by national institutions, in particular
the institutions participating in the {\it Gaia} Multilateral Agreement.

%%%%%%%%%%%%%%%%%%%%%%%%%%%%%%%%%%%%%%%%%%%%%%%%%%%%%%%%%%%%%%%%%%%%%%%%%%%%%%%%%%%%%%%%%%%

\begin{deluxetable}{l|ccccc}
\tablecolumns{5}  
\tablewidth{0pt} 
\tabletypesize{\tiny}
\setlength{\tabcolsep}{0.05in}
\tablenum{1} %\\               
\tablecaption{Relative astrometry and photometry}   
\tablehead{
\colhead{Property} & \colhead{A} & \colhead{B} & \colhead{C} & \colhead{D} & \colhead{G}\\
\colhead{} & \colhead{(S1)} & \colhead{(M2)} & \colhead{(M1)} & \colhead{(S2)} & \colhead{}\\
}
\startdata

\tableline
\multicolumn{4}{l}{\tiny {Measurements}}\\
\tableline

g & 15.60$\pm$0.01 &  15.72$\pm$0.01 &  16.45$\pm$0.02 & 18.09$\pm$0.01 & $\cdots$ \\
r & 15.40$\pm$0.01 &  15.55$\pm$0.01 &  16.21$\pm$0.01 & 17.74$\pm$0.01 & $\cdots$ \\
i & 15.36$\pm$0.01 &  15.57$\pm$0.02 &  16.15$\pm$0.02 & 17.74$\pm$0.02 & 19.50$\pm$0.20 \\
z & 15.23$\pm$0.03 &  15.50$\pm$0.05 &  16.02$\pm$0.01 & 17.68$\pm$0.03 & 18.95$\pm$0.13 \\
y & 14.92$\pm$0.01 &  15.23$\pm$0.02 & 15.76$\pm$0.02 & 17.36$\pm$0.02 & 19.20$\pm$0.24 \\
%$\Delta$i & 0.0 & 0.21 & 0.79 & 2.38 & 4.14 \\
$\Delta\alpha$cos($\delta$) & 0.000$\pm$0.004 & -1.185$\pm$0.004 & 1.271$\pm$0.005 & 0.410$\pm$0.004 & 0.240$\pm$0.050 \\
$\Delta\delta$ & 0.000$\pm$0.004 & -0.441$\pm$0.004 & -0.074$\pm$0.004 & -3.310$\pm$ 0.004 & -2.310$\pm$0.025 \\
%photoz & 0.818$^{+0.007}_{-0.009}$ & 0.732$^{+0.007}_{-0.006}$ & 0.912$^{+0.008}_{-0.009}$ & 0.914$^{+0.007}_{-0.009}$  \\

\tableline
\multicolumn{3}{l}{\tiny {Model Prediction}}\\
\tableline

$\Delta$m & 0.0 &  0.580$^{+0.17}_{-0.020}$ & 0.626$^{+0.005}_{-0.009}$ & 3.55$\pm$0.06 \\

\enddata

\tablecomments{
The first four columns are the four quasar images and the fifth column is the lens.
We also label the images according to the arrival time.
The first five rows are the magnitude measurements derived using Hostlens, corrected for extinction. 
The next two rows are the relative positions of each object relative to A. We assume a pixel scale of $0.256\arcsec$. $\Delta\alpha$ is positive toward the west, and $\Delta\delta$ toward the north.
%The following three rows are relative i magnitudes and relative positions in arcseconds averaged between the i and y bands. 
%The next row shows the photometric redshift estimates following \citet{wu10}.
The last row shows the model prediction for the relative magnitudes relative to image A, from the best-fit mass model.
}
\end{deluxetable}

\end{document}